\documentclass[sigconf]{acmart}

\usepackage{enumitem}
\usepackage{multirow}
\usepackage{tcolorbox}
\usepackage{pifont}
\newcommand{\cmark}{\ding{51}}
\AtBeginDocument{%
  }

\copyrightyear{2026}
\acmYear{2026}
\setcopyright{cc}
\setcctype{by}
\acmConference[KDD '26]{Proceedings of the 32nd ACM SIGKDD Conference on Knowledge Discovery and Data Mining V.2}{August 09--13, 2026}{Jeju Island, Republic of Korea}
\acmBooktitle{Proceedings of the 32nd ACM SIGKDD Conference on Knowledge Discovery and Data Mining V.2 (KDD '26), August 09--13, 2026, Jeju Island, Republic of Korea}
\acmDOI{10.1145/3770855.3817975}
\acmISBN{979-8-4007-2259-2/2026/08}

\begin{document}

\title{Uncertainty-aware Generative Recommendation}


\author{Chenxiao Fan}
\email{simonfan@mail.ustc.edu.cn}
\orcid{0009-0009-2509-7092}
\affiliation{
  \institution{University of Science and Technology of China}
  \city{Hefei}
  \country{China}
}

\author{Chongming Gao}
\authornote{Corresponding authors.}
\email{chongming.gao@gmail.com}
\orcid{0000-0002-5187-9196}
\affiliation{
  \institution{University of Science and Technology of China}
  \city{Hefei}
  \country{China}
}

\author{Yaxin Gong}
\email{gyx2022@mail.ustc.edu.cn}
\orcid{0009-0006-0321-2342}
\affiliation{
  \institution{University of Science and Technology of China}
  \city{Hefei}
  \country{China}
}

\author{Haoyan Liu}
\email{liuhaoyan@ustc.edu.cn}
\orcid{0000-0002-3249-2164}
\affiliation{
  \institution{University of Science and Technology of China}
  \city{Hefei}
  \country{China}
}

\author{Fuli Feng}
\email{fulifeng93@gmail.com}
\orcid{0000-0002-5828-9842}
\affiliation{
  \institution{University of Science and Technology of China}
  \city{Hefei}
  \country{China}
}

\author{Xiangnan He}
\authornotemark[1]
\email{xiangnanhe@gmail.com}
\orcid{0000-0001-8472-7992}
\affiliation{
  \institution{University of Science and Technology of China}
  \city{Hefei}
  \country{China}
}








\renewcommand{\shortauthors}{Chenxiao Fan et al.}

\begin{abstract}
Generative Recommendation has emerged as a transformative paradigm, reformulating recommendation as an end-to-end autoregressive sequence generation task.
Despite its promise, existing preference optimization methods typically rely on binary outcome correctness, suffering from a systemic limitation we term \textit{uncertainty blindness}.
This issue manifests in the neglect of the model's intrinsic generation confidence, the variation in sample learning difficulty, and the lack of explicit confidence expression, directly leading to unstable training dynamics and unquantifiable decision risks.
In this paper, we propose \textbf{Uncertainty-aware Generative Recommendation (UGR)}, a unified framework that leverages uncertainty as a critical signal for adaptive optimization.
UGR synergizes three mechanisms: (1) an uncertainty-weighted reward to penalize confident errors; (2) difficulty-aware optimization dynamics to prevent premature convergence; and (3) explicit confidence alignment to empower the model with confidence expression capabilities.
Extensive experiments demonstrate that UGR not only yields superior recommendation performance but also fundamentally stabilizes training, preventing the performance degradation often observed in standard methods. Furthermore, the learned confidence enables reliable downstream risk-aware applications.
Our project repository is available at: \texttt{https://github.com/cxfann/UGR}.
\end{abstract}




\maketitle

\newcommand\kddavailabilityurl{https://doi.org/10.5281/zenodo.20287710}
\ifdefempty{\kddavailabilityurl}{}{
\begingroup\small\noindent\raggedright\textbf{Resource Availability:}\\
The source code of this paper has been made publicly available at \url{\kddavailabilityurl}.
\endgroup
}

\begin{figure}[h]
  \centering
  \includegraphics[width=\linewidth]{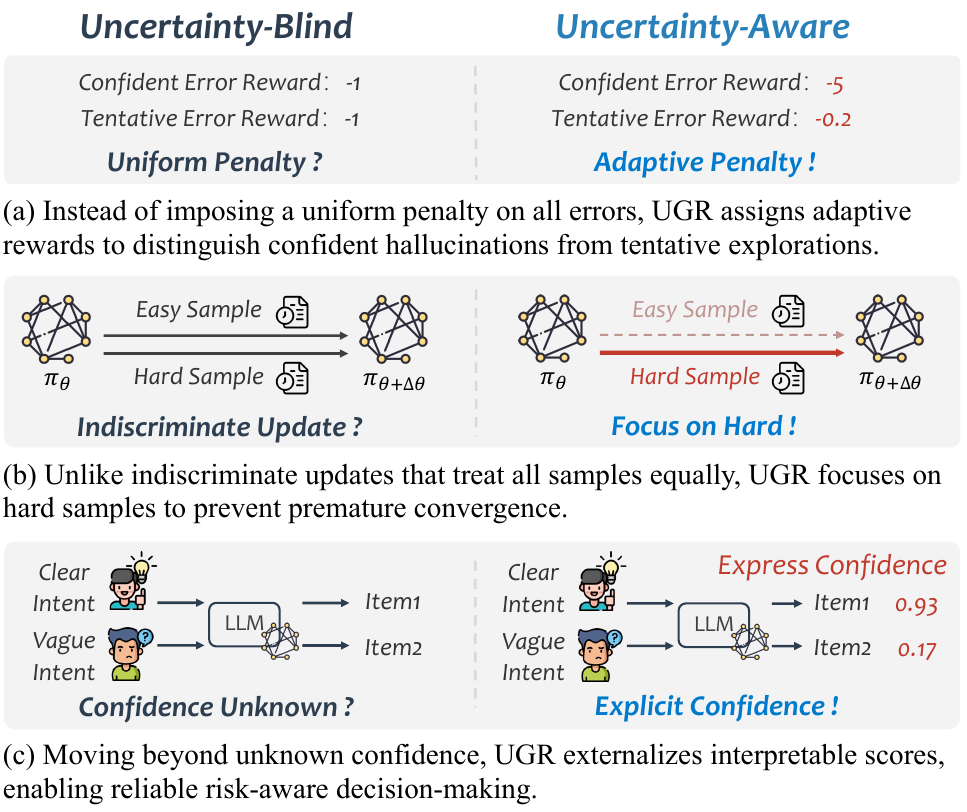}
  \caption{Comparison between existing Uncertainty-Blind methods (Left) and our proposed UGR (Right).}
  \label{fig:intro}
\end{figure}

\section{Introduction}
The field of recommender systems is undergoing a fundamental paradigm shift from discriminative ranking to generative modeling~\cite{hou2025generative,li2025survey,wang2025generative}.
Diverging from traditional discriminative approaches that rely on multi-stage ``retrieval-then-ranking'' pipelines to estimate user-item relevance scores~\cite{he2017neural,caser}, Generative Recommendation unifies the task as an end-to-end conditional sequence generation problem~\cite{tiger,rere}.
Under this paradigm, items are no longer treated as atomic indices but are serialized into sequences of discrete tokens, typically represented as Semantic IDs (SIDs)~\cite{lcrec,tiger,letter} or textual identifiers~\cite{geng2022recommendation,bigrec,sprec}.
Consequently, the underlying logic of recommendation is fundamentally transformed into an autoregressive decoding process, where the model constructs target items by sequentially sampling from a predicted probability distribution over the vocabulary~\cite{tallrec,bao2024decoding,lin2025order}.

To align the generative policy with user intent, preference optimization has been widely adopted in generative recommendation~\cite{liao2024rosepo,rlhf,dpo,cai2025mgfrec,cai2025korder,zhang2025unpaired,chen2026blade}.
However, existing objectives are typically predicated solely on outcome correctness~\cite{rere,minionerec}. This paradigm reduces the generation process to a binary right-or-wrong assessment, overlooking the variations in the model's internal confidence and sample learning difficulty~\cite{guo2017calibration,kadavath2022language}.
We term this systemic oversight \textit{uncertainty blindness} (illustrated in Figure~\ref{fig:intro}), which manifests in three critical limitations:
First, by imposing uniform penalties on all errors, the model fails to distinguish the severity of mistakes—specifically, differentiating between confident hallucinations characterized by high probability assignments to incorrect items, and tentative explorations stemming from a flat probability distribution.
Second, treating all training instances equally ignores the disparity in learning difficulty, leading to premature convergence on trivial samples while under-fitting hard ones;
Finally, the lack of explicit confidence expression prevents the model from quantifying decision risk, forcing it to rely solely on sampling strategies even in high-uncertainty scenarios.
Taken together, these issues underscore the necessity of explicitly modeling and leveraging uncertainty within the preference optimization loop.

To this end, rather than overlooking this intrinsic information, we propose to transform uncertainty into a high-value signal guiding model optimization.
We propose \textbf{Uncertainty-aware Generative Recommendation (UGR)}, a framework designed to explicitly model and leverage generative uncertainty to enhance robustness and trustworthiness.
Specifically, the framework incorporates three synergistic mechanisms:
First, we introduce an \textit{uncertainty-weighted reward}. Leveraging the model's intrinsic predictive distribution to distinguish the nature of errors, this mechanism imposes severe penalties on confident hallucinations while preserving the model's latitude for tentative exploration.
Second, we design \textit{difficulty-aware optimization dynamics}. By dynamically sensing the learning difficulty of each instance, we adaptively re-weight the optimization process, thereby preventing premature convergence on trivial patterns and ensuring persistent engagement with hard-to-learn samples.
Finally, we propose \textit{explicit confidence alignment}, which empowers the model to articulate confidence scores alongside recommendations, externalizing its internal certainty into quantifiable risk signals to support reliable decision-making.
Together, these mechanisms elevate the optimization objective from simple outcome fitting to robust uncertainty-aware learning.

Extensive experiments validate the effectiveness of UGR, revealing several critical and counter-intuitive findings.
Notably, we observe that standard preference optimization, lacking uncertainty awareness, often underperforms the Supervised Fine-Tuning (SFT) baseline. This degradation arises because coarse rewards treat all errors equally, introducing noisy signals when the model is uncertain.
In contrast, analyzing the training dynamics reveals that UGR stabilizes the learning trajectory: it prevents premature convergence on trivial samples while sustaining effective exploration.
Moreover, the aligned confidence serves not only as a regularization mechanism to boost ranking accuracy but also provides reliable signals for downstream risk-aware applications, such as list truncation and reranking.

The main contributions of this work are summarized as follows:
\begin{itemize}[leftmargin=*]
    \item We identify the issue of \textit{uncertainty blindness} in current preference optimization. We analyze how neglecting uncertainty undermines training stability and limits the model's capability.

    \item We propose \textbf{Uncertainty-aware Generative Recommendation (UGR)}, a unified framework that leverages uncertainty for adaptive optimization. Through joint training, UGR simultaneously optimizes recommendation quality and endows the model with explicit confidence expression.

    \item Extensive experiments demonstrate that UGR yields superior performance and fundamentally stabilizes training dynamics, preventing the degradation observed in standard methods. Furthermore, the learned confidence enables reliable downstream risk-aware applications.
\end{itemize}

Taken together, our work suggests that in generative recommendation, modeling uncertainty is not merely a calibration task, but an indispensable component of the preference optimization loop itself.

\begin{figure*}
  \centering
  \includegraphics[width=\linewidth]{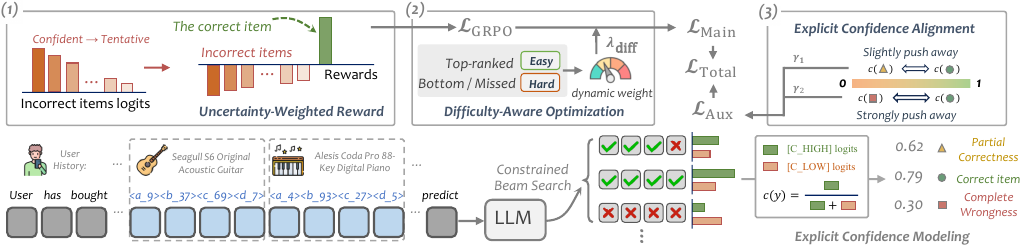}
  \caption{Overview of the proposed UGR framework.
  Given the user history, the model generates a candidate set via constrained beam search.
  These candidates effectively drive three synergistic mechanisms:
  (1) Uncertainty-weighted reward penalizes confident hallucinations based on logit intensity;
  (2) Difficulty-aware optimization adaptively re-weights the gradient budget based on ranking difficulty;
  (3) Explicit confidence alignment externalizes internal certainty into quantifiable risk signals.}
  \label{fig:model}
\end{figure*}

\section{Preliminary}
In this section, we explicitly formulate the generative recommendation task and clarify the distinction between implicit generation uncertainty and explicit confidence, laying the theoretical groundwork for our proposed method.

\subsection{Generative Recommendation Formulation}
Let $\mathcal{U}$ and $\mathcal{I}$ denote the sets of users and items, respectively.
For a user $u \in \mathcal{U}$, we represent their historical interactions as a time-ordered sequence
$H_u = [i_1, i_2, \dots, i_k]$, where each $i_j \in \mathcal{I}$ denotes an interacted item.

In the generative recommendation paradigm, the recommendation task is formulated as a conditional generation problem.
Given the user context $x$ derived from $H_u$, the model aims to generate an output sequence $y$ that represents the recommended item.
This process is modeled by a parameterized generative policy $\pi_\theta(y \mid x)$, which produces the output in an autoregressive manner:
\begin{equation}
\pi_\theta(y \mid x) = \prod_{j=1}^{L} P(t_j \mid x, y_{<j}; \theta),
\end{equation}
where $t_j$ denotes the $j$-th generated token and $L$ is the sequence length.

The final recommendation is obtained by decoding from the model's generative distribution.
In practice, each item may be represented by a sequence of discrete tokens from the vocabulary. In this work, we specifically consider SIDs, where the token sequence encodes the categorical hierarchy of the item.

To align the generative policy with user preference, the model is typically optimized to maximize a reward function $R(x, y)$ that evaluates the quality of the generated hypothesis against the ground truth $y^*$.
Standard approaches predominantly employ outcome-based rewards~\cite{sprec}.
The most common form is the binary reward, which strictly penalizes any mismatch:
\begin{equation}
R_{\text{bin}}(y, y^*) = \mathbb{I}(y = y^*).
\end{equation}
To address the uniform penalty applied to all incorrect generations by the binary reward, ranking-based rewards have been introduced~\cite{chen2024softmax,rere,minionerec}.
This paradigm aims to differentiate error severity by imposing stricter penalties on hard negatives—incorrect items assigned high rankings by the model.
Formally, this can be formulated as a rank-aware penalty:
\begin{equation}
R_{\text{rank}}(y, y^*) = \mathbb{I}(y = y^*) -  \frac{\mathbb{I}(y \neq y^*)}{\log(\text{rank}(y) + 1)},
\end{equation}
where $\text{rank}(y)$ denotes the ranking position of the generated item $y$ within the model's probability distribution.

However, both identity or rank paradigms treat the reward as a static signal derived solely from discrete outcomes, neglecting the model's intrinsic generation uncertainty and the varying difficulty of training instances.

\subsection{Confidence and Uncertainty in Recommendation}
\label{preliminary:uncertainty}
Generative recommendation fundamentally differs from factoid QA or logical reasoning tasks, which typically possess a unique, objectively correct ground truth.
In recommendation scenarios, user interaction is inherently subjective and stochastic. Consequently, uncertainty stems not merely from model capability limitations but serves as an intrinsic property characterizing the diversity of user interests and the distribution of potential behaviors~\cite{kendall2017uncertainties}.
We explicitly distinguish two forms of uncertainty:

\subsubsection{Implicit Generation Uncertainty.}
Mainstream LLMs generate recommendations via autoregressive decoding, where uncertainty is implicitly encoded in the probability distribution over the vocabulary~\cite{kweon2025uncertainty}.
Formally, for a generated recommendation $y$ under context $x$, implicit uncertainty $\mathcal{U}_{\text{imp}}(y \mid x)$ is often characterized by the negative log-likelihood of the generated sequence~\cite{xiao2019quantifying}:
\begin{equation}
\mathcal{U}_{\text{imp}}(y \mid x) = -\frac{1}{|y|} \sum_{t=1}^{|y|} \log P(y_t \mid y_{<t}, x).
\end{equation}
where $y_t$ denotes the token generated at step $t$, and $y_{<t}$ represents the prefix sequence. A higher $\mathcal{U}_{\text{imp}}$ indicates lower prediction probability, reflecting greater uncertainty.
Such implicit uncertainty provides a fine-grained reflection of the model's internal generation dynamics, capturing relative preference and ambiguity at the token level.
It is particularly effective for monitoring generation states (e.g., distinguishing determination from hesitation) and assessing the relative difficulty of different inputs.
However, tied to token-level objectives, implicit uncertainty is not explicitly designed to convey decision-level confidence or to directly support risk-aware recommendation.
Moreover, due to the normalization of probability distributions, implicit uncertainty limits its absolute comparability across samples, constraining its expressiveness in highly ambiguous scenarios.

\subsubsection{Explicit Confidence.}
Distinct from implicit statistics, explicit confidence $\mathcal{C}_{\text{exp}}(y)$ refers to the model's direct assessment of the overall quality or plausibility of a generated recommendation, typically articulated through dedicated output heads or confidence tokens~\cite{lin2022teaching,rlcr}.
In logic-intensive tasks, confidence learning often targets absolute calibration, aiming to align confidence values with empirical accuracy~\cite{guo2017calibration}.
In recommendation settings, however, such absolute calibration is less meaningful due to the extreme data sparsity and the generally low true click-through rate, which would force calibrated scores towards zero.
Instead, the relative ordering of confidence across different recommendations more naturally reflects the model's comparative belief in alternative decisions.
This ranking-based confidence complements implicit uncertainty by providing a decision-level signal aligned with recommendation objectives, facilitating robust preference modeling and risk-aware applications.

\section{Method}

\subsection{Overview}
In this section, we introduce the \textbf{Uncertainty-aware Generative Recommendation (UGR)}, designed to address the pervasive issue of uncertainty blindness in generative recommendation.
Diverging from traditional preference optimization methods that focus solely on outcome correctness, UGR reformulates preference alignment as an uncertainty-aware closed-loop system. As illustrated in Figure~\ref{fig:model}, the framework comprises three synergistic components.

To ensure that uncertainty signals are statistically measurable and comparable within this loop, UGR is grounded in two structural prerequisites.
First, by representing items as SIDs~\cite{tiger}, we leverage their fixed-length discrete structure to eliminate length-induced bias, rendering logits directly comparable across sequences while naturally preserving fine-grained partial correctness signals.
Second, we employ constrained beam search during rollout to generate a set of distinct candidates confined to the high-likelihood space~\cite{rere}. This ensures that logit discrepancies reflect the model's intrinsic hesitation regarding hard negatives, rather than stochastic noise or redundant sampling.

Building upon this solid foundation, UGR injects uncertainty awareness through three core mechanisms:
\begin{itemize}[leftmargin=*]
\item \textbf{Uncertainty-aware Rollout Reward:} Addressing the limitation of uniform penalties, this mechanism distinguishes the nature of errors during the rollout phase. By analyzing logit intensity, we impose stricter penalties on confident hallucinations, thereby providing denoised, high-quality signals for the subsequent optimization.
\item \textbf{Difficulty-aware Optimization Dynamics:} Recognizing the variation in learning difficulty, this module dynamically modulates the gradient budget based on instance-level generation difficulty. This prevents overfitting on trivial patterns while ensuring sustained and effective exploration for hard-to-learn samples.
\item \textbf{Explicit Confidence Alignment:} Leveraging the hierarchical structure of SIDs, this component aligns the model's explicit confidence with generation quality (i.e., complete vs. partial correctness). This endows the model with interpretable risk-aware capabilities, transforming internal uncertainty into actionable decision signals.
\end{itemize}

\subsection{Uncertainty-aware Rollout Reward}
Existing reward mechanisms are limited by their coarse-grained characterization of error severity. While basic binary rewards impose uniform penalties, improved ranking-based rewards, despite incorporating ordinal information, remain constrained by a rigid penalty scale. This discrete ranking signal obscures the model's intrinsic probability landscape: it fails to differentiate between errors stemming from highly confident hallucinations versus low-confidence hesitations, assigning identical penalties solely based on rank positions.

To break the bottleneck, we propose a transition from position-aware to distribution-aware optimization, leveraging the logit intensity of rollouts to construct a continuous uncertainty-aware reward.
Specifically, given a context $x$, we employ constrained beam search to sample a candidate set $\mathcal{G} = \{y_1, \dots, y_G\}$. We define the cumulative logit intensity of a sequence as $s(y) = \sum_{t} \log P(y_t | y_{<t}, x)$, and design a reward $r(y_i)$ based on $s(y)$ to adaptively penalize the set of incorrect generations $\mathcal{N}= \mathcal{G} \setminus \{y^*\}$, where $y^*$ denotes the ground-truth item sequence.
The proposed reward $r(y_i)$ is as below:
\begin{equation}
r(y_i) =
\begin{cases}
1.0, & \text{if } y_i = y^* , \\
- \frac{\exp(s(y_i))}{\sum_{y_j \in \mathcal{N}} \exp(s(y_j))}, & \text{if } y_i \in \mathcal{N},
\end{cases}
\label{eq:uncertainty_reward}
\end{equation}
This design augments ranking-based comparison with continuous modeling of error intensity. Intuitively, the higher the model's certainty in a specific error, the larger the magnitude of the corresponding negative reward.
Finally, adhering to the Group Relative Policy Optimization (GRPO) framework~\cite{grpo}, we normalize these rewards within each sampled group to compute the relative advantage:
\begin{equation}
\hat{A}_{i, t} = \frac{r(y_i) - \mu_r}{\sigma_r + \epsilon},
\label{equ:advantage}
\end{equation}
where $\mu_r$ and $\sigma_r$ denote the mean and standard deviation of rewards $\{r(y_j)\}_{j=1}^G$ in the current group, and $\epsilon$ is a small constant for numerical stability. This normalized advantage $\hat{A}_{i,t}$ serves as the input for the subsequent optimization objective.

\subsection{Difficulty-aware Optimization Dynamics}
While the uncertainty-aware rollout reward addresses intra-sample error severity, generative recommendation still suffers from significant inter-sample imbalance. User contexts vary drastically in inference difficulty, yet standard training paradigms typically optimize a uniform objective across all instances. This approach inherently overlooks the disparity in learning difficulty~\cite{lin2017focal}, often causing the model to overfit trivial patterns while under-exploring hard samples that contain rich structural information.
In generative recommendation, learning difficulty directly reflects generation uncertainty, since a sample is difficult precisely when the model cannot confidently locate the correct item.

To address this, we propose difficulty-aware optimization dynamics, which incorporates an adaptive re-weighting mechanism to shift the learning focus toward harder samples.
Since sample difficulty lacks explicit supervision, we adopt the ground truth's ranking position within the beam search candidates as a lightweight proxy, which effectively reflects the model's current capability to locate the target item.

Specifically, for a context $x$ and ground-truth $y^*$, let $\text{rank}(y^*)$ denote the position of $y^*$ within the candidate set $\mathcal{G}$ sorted by logits (0-indexed).
If $y^*$ is missed by the beam search ($y^* \notin \mathcal{G}$), indicating a high-uncertainty region, we assign the maximum difficulty $\text{rank}(y^*) = G - 1$.
The dynamic weight $\lambda_{\text{diff}}(x)$ is constructed as:
\begin{equation}
\lambda_{\text{diff}}(x) = \alpha + (1 - \alpha) \cdot \frac{\text{rank}(y^*)}{G - 1},
\label{eq:dynamic_weight}
\end{equation}
where $\alpha \in [0, 1]$ serves as a hyperparameter to ensure basic learning stability for easy samples (i.e., the lower bound of $\lambda_{\text{diff}}$ when $\text{rank}(y^*)=0$).

We integrate this adaptive weighting mechanism into the GRPO framework.
Standard GRPO optimizes the policy $\pi_\theta$ by maximizing the expected advantage of sampled generations. Based on the normalized advantage $\hat{A}_{i, t}$ derived in Eq.~(\ref{equ:advantage}), the unweighted objective for a single context $x$ is formulated as:
\begin{equation}
\mathcal{L}_{\text{GRPO}}(x; \theta) = - \frac{1}{G} \sum_{i=1}^{G} \frac{1}{|y_i|} \sum_{t=1}^{|y_i|} \frac{\pi_\theta(y_{i,t} \mid x, y_{i, <t})}{\pi_{\theta_{\text{old}}}(y_{i,t} \mid x, y_{i, <t})} \hat{A}_{i, t},
\label{eq:grpo_base}
\end{equation}
In practice, this objective is typically augmented with clipping or KL-regularization constraints to stabilize updates; we present the simplified form here to highlight the core optimization logic.

Finally, we propose the difficulty-aware instance loss by modulating the standard objective with the derived difficulty weight:
\begin{equation}
\mathcal{L}_{\text{Main}}(x; \theta) = \lambda_{\text{diff}}(x) \cdot \mathcal{L}_{\text{GRPO}}(x; \theta).
\label{eq:weighted_instance_loss}
\end{equation}
This formulation effectively propagates uncertainty awareness from the reward level (via $\hat{A}_{i, t}$) to the optimization dynamics level (via $\lambda_{\text{diff}}$). It allows the model to automatically balance the consolidation of simple patterns with the aggressive exploration of hard samples, laying a robust foundation for the subsequent confidence alignment task.

\subsection{Explicit Confidence Modeling and Alignment}

While the aforementioned modules effectively optimize training dynamics, the utilized uncertainty remains implicit and inaccessible for inference.
To bridge this, we propose explicit confidence alignment. This module serves a dual purpose: it externalizes intrinsic certainty into interpretable signals for risk-aware decision-making, and acts as semantic regularization to reinforce the model's comprehension of the hierarchical SID structure.

\paragraph{Verbalized Confidence Tokenization.}
We expand the vocabulary with a dedicated set of tokens: $V_{\text{conf}} = \{\texttt{[C\_HIGH]}, \texttt{[C\_LOW]}\}$.
Upon generating a recommendation sequence $y$ given context $x$, we perform a lightweight forward pass using an ``Append-and-Query'' strategy.
The explicit confidence score $c(y)$ is defined as the normalized probability of the high-confidence token:
\begin{equation}
c(y) =
\frac{P(\texttt{[C\_HIGH]} \mid x, y)}
{P(\texttt{[C\_HIGH]} \mid x, y) + P(\texttt{[C\_LOW]} \mid x, y)}.
\label{equ:conf}
\end{equation}
This formulation renders the confidence score a fully differentiable internal representation that updates synchronously with the generation policy.

\paragraph{Hierarchical Error Taxonomy.}
To provide high-quality alignment signals, we leverage the inherent tree-structured semantics of SIDs to structure the negative sample space into two distinct types:
\begin{itemize}[leftmargin=*]
    \item \textbf{Partial Correctness ($\mathcal{N}_{\text{partial}}$):}
    The generated sequence $y$ shares a non-empty prefix with the ground truth $y^*$ but deviates at the fine-grained item level. These samples typically correspond to hard negatives within the high-likelihood space.
    \item \textbf{Complete Wrongness ($\mathcal{N}_{\text{wrong}}$):}
    The sequence $y$ mismatches $y^*$ at the root level, indicating a significant semantic drift.
\end{itemize}

\paragraph{Hierarchical Confidence Alignment.}
Based on this taxonomy, we construct a multi-granular pair-wise ranking loss to map the semantic hierarchy onto the confidence space:
\begin{equation}
\mathcal{L}_{\text{Aux}} =
\gamma_1 \cdot \mathcal{L}_{\text{rank}}(y^*, \mathcal{N}_{\text{partial}})
+ \gamma_2 \cdot \mathcal{L}_{\text{rank}}(y^*, \mathcal{N}_{\text{wrong}}),
\label{eq:aux_loss}
\end{equation}
where $\mathcal{L}_{\text{rank}}$ adopts the logistic margin form:
\begin{equation}
\mathcal{L}_{\text{rank}}(y^*, S)
= \frac{1}{|S|} \sum_{y \in S}
\log \left( 1 + \exp \big( c(y) - c(y^*) \big) \right).
\end{equation}
By setting $\gamma_2 > \gamma_1$, we impose a stronger gradient penalty on complete errors, forcing their margin from the ground truth in the probability space to be significantly larger than that of partially correct samples.
This mechanism implicitly induces a confidence hierarchy,
$c(y^*) > c(y_{\text{partial}}) > c(y_{\text{wrong}})$,
thereby endowing the model with the ability to distinguish among different types of negative samples.

\subsection{Overall Objective and Training}
\label{subsec:overall_training}

We model the preference alignment of Generative Recommendation as a unified multi-task optimization framework. Given that both tasks share the same context inputs and generation process, for a batch $\mathcal{B}$ in a single iteration, we define the total training objective as the direct summation of the two losses:
\begin{equation}
\mathcal{L}_{\text{Total}}(\theta) = \frac{1}{|\mathcal{B}|} \sum_{x \in \mathcal{B}} \left( \mathcal{L}_{\text{Main}}(x; \theta) + \mathcal{L}_{\text{Aux}}(x; \theta) \right).
\label{eq:total_loss}
\end{equation}
Here, $\mathcal{L}_{\text{Main}}(x; \theta)$ refers to the difficulty-aware GRPO loss (Eq. (\ref{eq:weighted_instance_loss})), optimizing generation correctness and diversity; $\mathcal{L}_{\text{Aux}}(x; \theta)$ corresponds to the confidence ranking loss derived from the same rollout samples (Eq. (\ref{eq:aux_loss})).
Since the auxiliary loss is intrinsically scale-compatible with the normalized advantage-based main loss, and internally balanced by $\gamma_1, \gamma_2$, we do not introduce an additional global hyperparameter, thereby maintaining the simplicity and robustness of the optimization framework.

To efficiently optimize this objective, UGR adopts an end-to-end training strategy.
In each training step, the workflow strictly follows three steps:
\begin{itemize}[leftmargin=*]
    \item \textbf{Step 1} (Joint Rollout \& Evaluation):
    The model executes constrained beam search under the current policy $\pi_\theta$ to generate a candidate set $\mathcal{G}$. Crucially, this single set is simultaneously utilized to compute the reward for the main task and to construct hierarchical negative pairs for the auxiliary task, ensuring consistent variance across objectives.
    \item \textbf{Step 2} (Dynamic Weighting):
    Based on the relative ranking position of the ground truth $y^*$ within $\mathcal{G}$, we compute the instance-level dynamic weight $\lambda_{\text{diff}}(x)$ to adaptively regulate the gradient contribution of $\mathcal{L}_{\text{Main}}$.
    \item \textbf{Step 3} (Unified Backpropagation):
    A forward pass is performed on Eq. (\ref{eq:total_loss}) to output both the likelihood of recommendation tokens and the distribution of confidence tokens. Finally, gradients from $\mathcal{L}_{\text{Main}}$ and $\mathcal{L}_{\text{Aux}}$ are aggregated to update the shared parameters $\theta$.
\end{itemize}

Through this joint optimization mechanism, UGR ensures that confidence modeling is deeply embedded in the policy learning dynamics, realizing the synergistic evolution of generative capability and risk awareness.

\begin{table*}[t]
\centering
\caption{Overall performance comparison on three datasets. The baselines are categorized into traditional models (SASRec), LLM-based models (covering text-based methods: BIGRec, SPRec; and SID-based generative methods: TIGER, MiniOneRec), and reasoning-enhanced models (ReaRec, R$^2$ec). The best results are highlighted in bold, and the second-best results are underlined.}
\label{tab:overall_performance}
\setlength{\tabcolsep}{1.5pt}
\begin{tabular}{l ccccc c ccccc c ccccc}
\toprule
\multirow{2}{*}{\textbf{Method}} & \multicolumn{5}{c}{\textbf{Office}} && \multicolumn{5}{c}{\textbf{Industrial}} && \multicolumn{5}{c}{\textbf{Yelp}} \\
\cmidrule{2-6} \cmidrule{8-12} \cmidrule{14-18}
& HR@1 & HR@5 & HR@10 & NG@5 & NG@10 && HR@1 & HR@5 & HR@10 & NG@5 & NG@10 && HR@1 & HR@5 & HR@10 & NG@5 & NG@10 \\
\midrule
SASRec      & 0.0176 & 0.0771 & 0.1028 & 0.0489 & 0.0572 && 0.0276 & 0.0732 & 0.0894 & 0.0498 & 0.0551 && 0.0059 & 0.0253 & 0.0441 & 0.0153 & 0.0213 \\
BIGRec       & 0.0714 & 0.1053 & 0.1198 & 0.0898 & 0.0944 && 0.0272 & 0.0907 & 0.1143 & 0.0635 & 0.0710 && 0.0036 & 0.0089 & 0.0119 & 0.0063 & 0.0073 \\
SPRec    & 0.0665 & 0.0885 & 0.1013 & 0.0787 & 0.0828 && 0.0305 & 0.0970 & 0.1244 & 0.0689 & 0.0780 && 0.0066 & 0.0096 & 0.0113 & 0.0081 & 0.0087 \\
TIGER       & 0.0715 & 0.1146 & 0.1361 & 0.0946 & 0.1015 && 0.0489 & 0.0954 & 0.1217 & 0.0744 & 0.0828 && 0.0057 & \textbf{0.0283} & 0.0460 & \underline{0.0167} & 0.0223 \\
MiniOneRec  & \underline{0.0823} & \underline{0.1314} & \underline{0.1479} & \underline{0.1084} & \underline{0.1137} && \underline{0.0745} & \underline{0.1080} & \underline{0.1276} & \underline{0.0917} & \underline{0.0980} && \underline{0.0087} & 0.0243 & 0.0420 & 0.0165 & 0.0226 \\
ReaRec      & 0.0782 & 0.1164 & 0.1380 & 0.0990 & 0.1060 && 0.0529 & 0.0952 & 0.1148 & 0.0760 & 0.0823 && 0.0074 & 0.0258 & \underline{0.0470} & 0.0163 & \underline{0.0232} \\
R$^2$ec        & 0.0496 & 0.1006 & 0.1364 & 0.0758 & 0.0873 && 0.0372 & 0.0886 & 0.1274 & 0.0635 & 0.0761 && 0.0066 & 0.0253 & 0.0464 & 0.0161 & 0.0227 \\
\midrule
\textbf{UGR} & \textbf{0.0904} & \textbf{0.1407} & \textbf{0.1598} & \textbf{0.1169} & \textbf{0.1231} && \textbf{0.0747} & \textbf{0.1179} & \textbf{0.1393} & \textbf{0.0965} & \textbf{0.1033} && \textbf{0.0096} & \underline{0.0281} & \textbf{0.0473} & \textbf{0.0188} & \textbf{0.0250} \\
\bottomrule
\end{tabular}
\end{table*}

\begin{table*}[t]
\centering
\caption{Ablation study of UGR on three datasets. (a) is the base model with SFT only. (b1)-(b3) denote variants using GRPO with different reward functions: 0/1 Reward, Rank-based Reward, and Uncertainty-Weighted Reward (UW-Reward). (c) adds Difficulty-aware Optimization upon (b3), and (d) is the full UGR model further incorporating Explicit Confidence Alignment. \textbf{Bold} indicates the best performance.}
\label{tab:ablation_study}
\setlength{\tabcolsep}{1.5pt}
\begin{tabular}{l cccc c cccc c cccc}
\toprule
\multirow{2}{*}{\textbf{Method / Variant}} & \multicolumn{4}{c}{\textbf{Office}} && \multicolumn{4}{c}{\textbf{Industrial}} && \multicolumn{4}{c}{\textbf{Yelp}} \\
\cmidrule{2-5} \cmidrule{7-10} \cmidrule{12-15}
& HR@5 & HR@10 & NG@5 & NG@10 && HR@5 & HR@10 & NG@5 & NG@10 && HR@5 & HR@10 & NG@5 & NG@10 \\
\midrule
(a) Only SFT & 0.1374 & 0.1563 & 0.1152 & 0.1213 && 0.1104 & 0.1378 & 0.0923 & 0.1010 && 0.0264 & 0.0423 & 0.0165 & 0.0217 \\
\midrule
(b1) + GRPO (0/1 Reward) & 0.1283 & 0.1485 & 0.1056 & 0.1120 && 0.1041 & 0.1303 & 0.0845 & 0.0931 && 0.0249 & 0.0405 & 0.0170 & 0.0220 \\
(b2) + GRPO (Rank Reward) & 0.1291 & 0.1468 & 0.1079 & 0.1135 && 0.1076 & 0.1321 & 0.0878 & 0.0958 && 0.0232 & 0.0423 & 0.0157 & 0.0219 \\
(b3) + GRPO (UW-Reward) & 0.1349 & 0.1522 & 0.1137 & 0.1193 && 0.1144 & 0.1338 & \textbf{0.0967} & 0.1029 && 0.0234 & 0.0424 & 0.0160 & 0.0222 \\
\midrule
(c) + Difficulty-Aware Opt.  & 0.1378 & 0.1579 & 0.1158 & 0.1223 && 0.1151 & 0.1349 & \textbf{0.0967} & 0.1030 && 0.0264 & 0.0454 & 0.0178 & 0.0239 \\
\midrule
(d) \textbf{UGR (Full)}& \textbf{0.1407} & \textbf{0.1598} & \textbf{0.1169} & \textbf{0.1231} && \textbf{0.1179} & \textbf{0.1393} & 0.0965 & \textbf{0.1033} && \textbf{0.0281} & \textbf{0.0473} & \textbf{0.0188} & \textbf{0.0250} \\
\bottomrule
\end{tabular}
\end{table*}

\section{Experiments}
In this section, we conduct experiments to answer the following
research questions (RQs):

\begin{itemize}[leftmargin=*]
    \item \textbf{RQ1:} How does UGR perform compared to existing baselines in generative recommendation tasks?
    \item \textbf{RQ2:} How does each uncertainty-aware component contribute to the overall effectiveness of the proposed framework?
    \item \textbf{RQ3:} Can UGR effectively stabilize the training process and mitigate the issue of uncertainty blindness?
    \item \textbf{RQ4:} Does the externalized confidence signal serve as a reliable indicator for downstream risk-aware applications?
\end{itemize}

\subsection{Experimental Setup}

\subsubsection{Datasets.}
We conduct experiments on three real-world benchmarks: Office Products (Office) and Industrial and Scientific (Industrial) from the Amazon Review Dataset\footnote{\url{https://jmcauley.ucsd.edu/data/amazon}}, along with the Yelp dataset\footnote{\url{https://www.yelp.com/dataset}}.
Following previous works~\cite{minionerec} we partition the data into training, validation, and test sets with an 8:1:1 ratio, strictly adhering to the chronological order of interactions.
Detailed preprocessing steps and dataset statistics are provided in Appendix~\ref{appendix:datasets}.

\subsubsection{Evaluation Setting and Metrics.}
We adopt two widely used metrics to evaluate top-$K$ recommendation performance: Hit Ratio (HR@$K$) and Normalized Discounted Cumulative Gain (NG@$K$), with $K \in \{1, 5, 10\}$.
During the inference phase, we employ beam search with a beam size of 10 to generate the ranking list.

\subsubsection{Baselines.}
We compare UGR against a comprehensive set of baselines categorized into three groups:
(1) Traditional Models, represented by SASRec~\cite{sasrec};
(2) LLM-based Models, which are further divided into two streams: text-based methods (BIGRec~\cite{bigrec}, SPRec~\cite{sprec}) that rely on grounding mechanisms, and SID-based generative methods (TIGER~\cite{tiger}, MiniOneRec~\cite{minionerec}) that utilize SID for direct generation; and
(3) Reasoning-enhanced Models, including ReaRec~\cite{rearec} and R$^2$ec~\cite{rrec}.
Further details are provided in Appendix~\ref{appendix:baselines}.

\subsubsection{Implementations.}
All experiments are conducted on 8 NVIDIA A100 GPUs.
We utilize Qwen3-8B~\cite{qwen3} as the backbone LLM, representing items via 4-layer SIDs.
The training pipeline consists of two stages:
First, we perform SFT for up to 10 epochs with early stopping.
Second, initialized from the SFT checkpoint, we conduct the proposed preference optimization as described in Section~\ref{subsec:overall_training} for 2 epochs.
Crucially, during the training rollout phase, we utilize beam search with a group size of $G=16$ to construct candidate sets for reward calculation and alignment.
Further implementation details are available in Appendix~\ref{appendix:implementation}.

\subsection{Overall Performance Comparison (RQ1)}
In this section, we present the comparative results of UGR against all baselines across three datasets. The performance metrics are summarized in Table~\ref{tab:overall_performance}.

The results reveal four key observations:
(1) UGR consistently achieves the best performance across all datasets and metrics, establishing a new state-of-the-art. The significant margin over the strongest baselines validates the robustness of our uncertainty-aware framework.
(2) SID-based generative methods (TIGER, MiniOneRec, UGR) generally surpass traditional (SASRec) and text-based (BIGRec) models, confirming that combining LLM semantics with structured representations captures complex preferences better than shallow embeddings or open-ended generation.
(3) Crucially, UGR significantly outperforms MiniOneRec. Given their shared backbone, this gain is strictly attributed to our resolution of uncertainty blindness, where weighted rewards and difficulty-aware dynamics provide superior optimization signals over binary feedback.
(4) UGR also outperforms reasoning-enhanced models (ReaRec, R$^2$ec), indicating that explicit uncertainty modeling offers a more direct and robust signal for ranking refinement.

\subsection{Component Analysis and Ablation (RQ2)}
\label{sec:ablation}
To evaluate the contribution of each component in UGR, we conduct a detailed ablation study in Table~\ref{tab:ablation_study}, incrementally introducing different reward functions and optimization dynamics starting from the SFT baseline.

\subsubsection{Impact of Uncertainty-Weighted Reward.}
First, we observe a critical phenomenon: standard preference optimization (Row b1-b2) fails to surpass the SFT baseline. This confirms our hypothesis of uncertainty blindness—coarse binary or ranking rewards introduce noisy signals by treating confident and tentative errors equally. In contrast, introducing the uncertainty-weighted reward (Row b3) significantly recovers performance, effectively matching SFT. This validates that distinguishing error severity based on generation confidence is a prerequisite for effective optimization. However, reward-level improvements alone are insufficient to significantly surpass SFT, necessitating advanced optimization dynamics.

\subsubsection{Effectiveness of Difficulty-aware Optimization.} Building upon (b3), incorporating difficulty-aware optimization (Row c) yields consistent gains across datasets. By dynamically reweighting instances based on difficulty and uncertainty, this strategy prevents overfitting to easy samples and sustains exploration, and its specific impact on training stability will be further analyzed in Sec.~\ref{rq3}.

\subsubsection{Benefit of Explicit Confidence Alignment.} Finally, the full UGR framework (Row d) achieves the best performance by jointly optimizing with explicit confidence alignment. Beyond enabling confidence expression, the auxiliary ranking loss acts as a semantic regularizer, reinforcing the model's structural understanding of SIDs to boost ranking accuracy by enforcing self-assessment. This module also enables the risk-aware applications explored in Sec.~\ref{rq4}.

Beyond the incremental ablation above, we further investigate the mutual synergy among the three modules through a complete decomposition over all seven combinations, and validate our binary confidence token design against alternatives. All analyses are deferred to Appendix~\ref{appendix:ablation}.

\begin{figure}
  \centering
  \includegraphics[width=\linewidth]{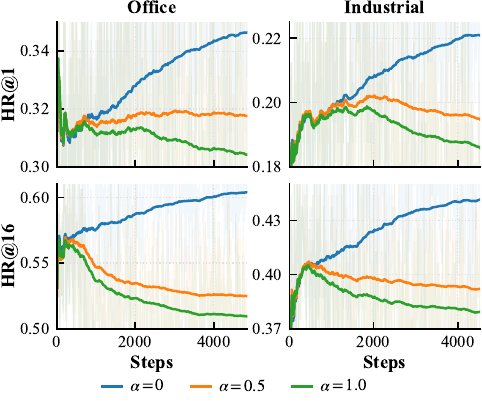}
  \caption{Training dynamics of HR@1 and HR@16 on Office and Industrial datasets under varying difficulty coefficients $\alpha$. Standard uniform weighting ($\alpha=1.0$) leads to performance collapse due to overfitting, whereas UGR's difficulty-aware optimization ($\alpha=0$) prevents degradation and enables sustained learning.}
  \label{fig:dynamics}
\end{figure}

\subsection{Training Dynamics Analysis (RQ3)}
\label{rq3}
To investigate the impact of difficulty-aware optimization on training stability and exploration capability, we visualize the training trajectories of HR@1 (ranking accuracy) and HR@16 (candidate coverage, aligned with the training beam width $G=16$) under varying difficulty coefficients $\alpha$ in Figure~\ref{fig:dynamics}.

\subsubsection{Instability of Uniform Optimization.}
Standard uniform weighting ($\alpha=1.0$) exhibits severe instability. As shown, it suffers from a distinct pattern of early convergence followed by regression. After an initial peak, both HR@1 and HR@16 degrade significantly. This indicates that the model overfits to trivial samples, causing the effective search space to shrink prematurely (dropping HR@16) and hampering the retrieval of hard items.

\subsubsection{Sustained Exploration via Difficulty Awareness.}
In contrast, UGR ($\alpha=0$) stabilizes training, maintaining a robust, monotonic upward trend. By suppressing gradients from easy samples, the optimization is compelled to persist in refining hard instances. This mechanism ensures sustained candidate coverage (High HR@16), which effectively translates diversity into precise top-tier recommendations (High HR@1), extending the effective learning horizon.

\begin{figure}
  \centering
  \includegraphics[width=\linewidth]{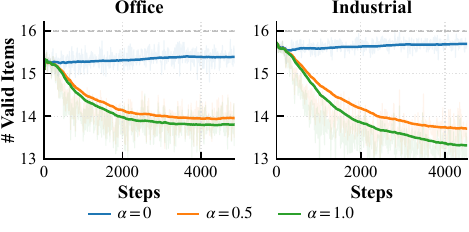}
  \caption{Average number of unique valid items per rollout (beam size $G{=}16$). As the standard variant ($\alpha{=}1.0$) becomes overconfident on easy samples, candidate diversity gradually erodes, whereas UGR ($\alpha{=}0$) sustains a near-saturated search space.}
  \label{fig:rollout_items}
\end{figure}

\subsubsection{Analysis of Candidate Collapse.}
Figure~\ref{fig:rollout_items} exposes the underlying failure mode of standard preference optimization, which we term \textit{candidate collapse}. The uniform variant ($\alpha{=}1.0$) exhibits a steady erosion in the number of unique valid items retrieved per rollout. This degradation stems from an \textit{overconfidence trap} on easy samples: the uniform objective forces the model to push the probability of trivial ground-truth tokens toward the extreme, yielding an excessively peaked distribution. As a result, beam search is left with progressively fewer alternative paths to explore, and the pool of valid candidates available for contrastive learning steadily shrinks.

In contrast, UGR ($\alpha{=}0$) resolves this pathology by dynamically down-weighting the gradients of easy samples, which directly suppresses such overconfidence and prevents the model from over-fitting to trivial patterns. By maintaining a reasonable rather than degenerate distribution, UGR preserves a healthy search space, ensuring that beam search consistently retrieves a diverse set of hard candidates to fuel stable optimization.

\begin{table}[t]
    \centering
    \caption{Performance comparison of Confidence-based Re-ranking on the Office dataset.}
    \label{tab:rerank}
    \resizebox{0.475\textwidth}{!}{
    \begin{tabular}{l c c c c}
        \toprule
        \textbf{Method} & \textbf{NDCG@1} & \textbf{NDCG@3} & \textbf{NDCG@5} & \textbf{NDCG@10} \\
        \midrule
        UGR (Original Rank)      & 0.0904 & 0.1102 & 0.1169 & 0.1231 \\
        UGR + Conf. Re-rank & \textbf{0.0933} & \textbf{0.1121} & \textbf{0.1185} & \textbf{0.1245} \\
        \midrule
        \textit{Gain ($\Delta$)} & \textit{+0.0029} & \textit{+0.0018} & \textit{+0.0016} & \textit{+0.0014} \\
        \bottomrule
    \end{tabular}
    }
\end{table}

\begin{figure}
  \centering
  \includegraphics[width=\linewidth]{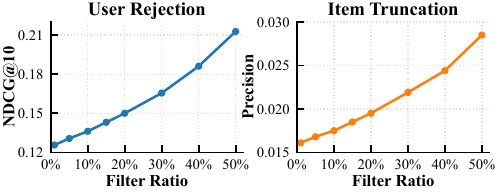}
  \caption{Performance of risk-aware applications on the Office dataset. User-level Rejection (Left): Filtering low-confidence requests steadily improves NDCG@10. Item-level Truncation (Right): Dynamically removing low-confidence tail items effectively enhances Precision.}
  \label{fig:risk_aware}
\end{figure}

\subsection{Utility of Explicit Confidence Signals (RQ4)}
\label{rq4}
Having established the training benefits in Section~\ref{sec:ablation}, we now focus on the practical utility of explicit confidence signals during the inference phase. Using the Office dataset, we evaluate their reliability as a decision basis across two dimensions: ranking enhancement and risk-aware recommendation.

\subsubsection{Confidence-based Ranking Enhancement}
First, we investigate the impact of incorporating confidence signals to re-rank candidates generated by beam search. As shown in Table~\ref{tab:rerank}, this integration yields consistent improvements across all metrics compared to using logits alone. This suggests that explicit confidence provides an additional assessment of logical correctness complementary to generation probability, effectively correcting high-confidence but factually incorrect hallucinations.

\subsubsection{Risk-aware Recommendation}
Beyond ranking accuracy, reliable recommendation requires effective risk control. We design two scenarios to verify the risk perception capability of confidence signals at different granularities (Figure~\ref{fig:risk_aware}):

\paragraph{User-level Rejection.}
Existing systems typically enforce mandatory output for all user requests. However, for hard samples where the model is extremely uncertain, such forced recommendations often result in a poor user experience. Figure~\ref{fig:risk_aware} (Left) shows that filtering requests based on confidence thresholds leads to a monotonic increase in NDCG@10. This validates that low confidence accurately identifies hard-to-model samples, enabling selective recommendation—rejecting risky queries to ensure system reliability.

\paragraph{Item-level Truncation.}
Simultaneously, at the item level, traditional Top-$K$ strategies enforce a fixed-length list (e.g., 10 items), which inevitably introduces noise or fillers at the tail.
Figure~\ref{fig:risk_aware} (Right) shows that when we dynamically truncate tail items based on confidence scores, the Precision of the recommendation list increases steadily.
This indicates that the confidence signal effectively identifies invalid items, supporting a shift from fixed-length to dynamic-length recommendation. This paradigm reduces invalid exposure to users while maintaining high accuracy.

\subsection{Hyperparameter Sensitivity}
\label{sec:sensitivity}

Table~\ref{tab:sensitivity} examines the sensitivity of the margin hyperparameters $\gamma_1$ (partial correctness) and $\gamma_2$ (complete wrongness) in Eq.~(\ref{eq:aux_loss}).

\begin{table}[t]
\centering
\caption{Sensitivity of margin hyperparameters $\gamma_1$ and $\gamma_2$.}
\label{tab:sensitivity}
\resizebox{\linewidth}{!}{
\begin{tabular}{cc cc cc cc}
\toprule
\multirow{2}{*}{$\gamma_2$} & \multirow{2}{*}{$\gamma_1$} & \multicolumn{2}{c}{\textbf{Office}} & \multicolumn{2}{c}{\textbf{Industrial}} & \multicolumn{2}{c}{\textbf{Yelp}} \\
\cmidrule(lr){3-4} \cmidrule(lr){5-6} \cmidrule(lr){7-8}
& & HR@10 & NG@10 & HR@10 & NG@10 & HR@10 & NG@10 \\
\midrule
\multirow{3}{*}{0.5}
& 0.0 & 0.1587 & 0.1227 & 0.1389 & 0.1032 & 0.0473 & 0.0250 \\
& 0.1 & 0.1598 & 0.1231 & 0.1393 & 0.1033 & 0.0457 & 0.0246 \\
& 0.5 & 0.1579 & 0.1221 & 0.1382 & 0.1030 & 0.0464 & 0.0243 \\
\midrule
\multirow{3}{*}{1.0}
& 0.0 & 0.1595 & 0.1226 & 0.1386 & 0.1034 & 0.0469 & 0.0249 \\
& 0.1 & 0.1600 & 0.1227 & 0.1393 & 0.1036 & 0.0466 & 0.0247 \\
& 0.5 & 0.1581 & 0.1217 & 0.1373 & 0.1029 & 0.0465 & 0.0244 \\
\bottomrule
\end{tabular}
}
\end{table}

Two observations emerge. First, all configurations yield comparable performance across datasets, indicating that UGR is not sensitive to the specific choice of these margins. Second, setting $\gamma_1$ to a small positive value (e.g., 0.1) while keeping $\gamma_2$ larger consistently achieves the best or near-best results. In contrast, when $\gamma_1$ approaches $\gamma_2$ (e.g., both at 0.5), performance degrades slightly, as the reward can no longer differentiate partially correct outputs from completely wrong ones. This validates our hierarchical margin design: assigning distinct penalty levels to different error severities provides a more informative learning signal than a uniform penalty.

\section{Related Work}
In this section, we review the literature relevant to our work from three perspectives: generative recommendation paradigms, preference alignment strategies, and uncertainty modeling in LLMs.

\subsection{Generative Recommendation}
Propelled by the remarkable capabilities of LLMs, recommender systems are undergoing a fundamental paradigm shift from discriminative ranking to generative modeling~\cite{wu2024survey,geng2022recommendation}.
Early explorations primarily leveraged the general text generation capabilities of LLMs to directly verbalize item titles~\cite{tallrec,zhang2025collm,wang2023enhancing}.
However, to address the validity challenges associated with unstructured generation, existing methods often resort to external grounding mechanisms~\cite{sprec,bigrec} or restrict the generation space to input candidate lists~\cite{liao2024rosepo}, which fundamentally constrains the end-to-end potential of generative recommendation.
With the advent of works such as TIGER~\cite{tiger} and LC-Rec~\cite{lcrec}, generative frameworks based on SIDs have gained prominence.
These approaches encode items into fixed-length discrete token sequences with hierarchical structures, and further enhance representational semantic density by incorporating collaborative information~\cite{letter}.
Recent studies have further explored process-supervised tuning, agentic feedback loops, proactive recommendation, and diversity-controllable generation for LLM-based recommenders~\cite{gao2025flow,cai2025agentic,wang2025tunable,chen2025dlcrec}.
Nevertheless, the current generative paradigm remains limited to simple item index prediction—merely sampling item IDs based on probability distributions—while lacking the capability to explicitly articulate confidence levels simultaneously with the recommendation output.
To bridge this gap, our proposed UGR endows the model with the ability to articulate confidence alongside recommendation generation, thereby enabling reliable risk-aware decision-making.

\subsection{Preference Alignment in Recommendation}
Solely relying on SFT is insufficient to fully align generative models with complex recommendation objectives~\cite{gao2026integrating}. To bridge the gap between next-token prediction and ranking optimization, preference alignment techniques have been widely adopted~\cite{bai2024aligning,chen2024softmax,kong2025think,cai2025korder,zhang2025unpaired,chen2026blade}.
Given the prohibitive computational costs of online reinforcement learning methods like PPO~\cite{ppo}, recent research has pivoted towards more efficient offline or group-wise optimization paradigms~\cite{lin2025rec,ding2026towards}.
RL-style adaptation has also been used for instance-wise prompt personalization~\cite{mao2025reinforcedprompt}.
Specifically, DPO~\cite{dpo} and its variants~\cite{sprec,liao2024rosepo} are extensively used for pairwise preference learning.
Meanwhile, GRPO~\cite{grpo}, distinguished by its critic-free design, has been employed by works such as ReRe~\cite{rere} and MiniOneRec~\cite{minionerec} to perform list-wise optimization on high-quality candidates sampled via beam search.
Furthermore, FLAME attempts to transcend binary reward limitations by introducing step-wise reward shaping~\cite{flame}.
However, these approaches remain fundamentally constrained by an outcome-correctness perspective. Relying solely on label correctness, they treat confident hallucinations and tentative explorations indistinguishably and ignore the disparity in learning difficulty across samples, falling into the trap of uncertainty blindness.
In contrast, UGR transforms uncertainty into high-value feedback signals, achieving adaptive optimization based on error nature and sample difficulty.

\subsection{Confidence and Uncertainty in LLMs and Recommendation}
Uncertainty estimation has become a pivotal research topic in the general LLM community~\cite{huang2025survey}.
Existing studies indicate that LLMs often exhibit significant confidence bias in complex tasks like code generation, lacking accurate self-assessment capabilities~\cite{tian2023just,kadavath2022language,code_calibration}.
To address this, works like RLCR introduce reinforcement learning with calibration rewards to align confidence with correctness, demonstrating that high-quality uncertainty signals can reciprocally enhance reasoning performance~\cite{rlcr}.
However, these approaches predominantly pursue absolute calibration~\cite{guo2017calibration}, aiming to strictly align predicted probabilities with objective accuracy.
As discussed in Sec.~\ref{preliminary:uncertainty}, the inherent subjectivity and sparsity of recommendation tasks render such absolute numerical values less meaningful~\cite{zhang2025vague}.
While some works analyze uncertainty in recommendation~\cite{kweon2025uncertainty}, its effective utilization in the generative paradigm remains unexplored.
Our work fills this gap by leveraging uncertainty signals to enhance the robustness of preference alignment, while simultaneously endowing the model with explicit risk-awareness capabilities.

\section{Conclusion}
In this paper, we identified \textit{uncertainty blindness} as a critical bottleneck in generative recommendation.
To bridge this gap, we proposed UGR, a unified framework that synergizes uncertainty-weighted rewards, difficulty-aware optimization, and explicit confidence alignment.
Our extensive experiments demonstrate that UGR not only establishes new state-of-the-art results but also effectively resolves the training instability issue common in standard methods.
Moreover, the learned confidence scores empower the model with reliable self-assessment capabilities for risk-aware decision-making.
While validated on three benchmarks of moderate scale, UGR operates on optimization dynamics and reward shaping rather than dataset-specific features, and we expect its principles to generalize to larger-scale settings.

Looking ahead, this work opens up several promising directions.
First, while validated within the GRPO framework, our core insight of uncertainty-aware optimization is generalizable. Future work will explore adapting this paradigm to other preference alignment strategies, such as DPO, to further improve sample efficiency.
Second, we aim to explore more fine-grained uncertainty decomposition mechanisms to provide richer interpretability for decision-making.
Finally, we plan to extend UGR to online interactive settings and dynamic domains, where rapidly shifting user interests amplify the difficulty variance that UGR is designed to handle. We seek to build a truly adaptive and risk-aware recommendation system.

\begin{acks}
This work is supported by the National Natural Science Foundation of China (62402470, U24B20180, 62525211), the Fundamental Research Funds for the Central Universities of China (WK2100000053), Anhui Provincial Natural Science Foundation (2408085QF189), This research is supported by the advanced computing resources provided by the Supercomputing Center of the USTC.
\end{acks}

\clearpage

\bibliographystyle{ACM-Reference-Format}
\bibliography{ref}

@inproceedings{sasrec,
  title={Self-attentive sequential recommendation},
  author={Kang, Wang-Cheng and McAuley, Julian},
  booktitle={2018 IEEE international conference on data mining (ICDM)},
  pages={197--206},
  year={2018},
  organization={IEEE}
}

@article{bigrec,
  title={A bi-step grounding paradigm for large language models in recommendation systems},
  author={Bao, Keqin and Zhang, Jizhi and Wang, Wenjie and Zhang, Yang and Yang, Zhengyi and Luo, Yanchen and Chen, Chong and Feng, Fuli and Tian, Qi},
  journal={ACM Transactions on Recommender Systems},
  volume={3},
  number={4},
  pages={1--27},
  year={2025},
  publisher={ACM New York, NY}
}

@article{gao2026integrating,
  title={Integrating Large Language Models with Reinforcement Learning: A Survey of {LLM-RL} Synergistic Recommendation},
  author={Gao, Mengyao and Gao, Chongming and Tang, Jiakai and Zhang, Jingsen and Zhao, Xinpeng and Wang, Bohao and Chen, Jiawei and He, Haoran and Pan, Ling and Chen, Xu and Xin, Xin and Cai, Qingpeng and Jiang, Peng and Gai, Kun and Liu, Haoyan and Feng, Fuli and He, Xiangnan},
  journal={TechRxiv},
  year={2026},
  doi={10.36227/techrxiv.177155631.17855475/v1},
  url={https://doi.org/10.36227/techrxiv.177155631.17855475/v1},
  note={Preprint}
}

@inproceedings{sprec,
  title={Sprec: Self-play to debias llm-based recommendation},
  author={Gao, Chongming and Chen, Ruijun and Yuan, Shuai and Huang, Kexin and Yu, Yuanqing and He, Xiangnan},
  booktitle={Proceedings of the ACM on Web Conference 2025},
  pages={5075--5084},
  year={2025},
  doi={10.1145/3696410.3714524}
}

@inproceedings{gao2025flow,
  title={Process-Supervised {LLM} Recommenders via Flow-guided Tuning},
  author={Gao, Chongming and Gao, Mengyao and Fan, Chenxiao and Yuan, Shuai and Shi, Wentao and He, Xiangnan},
  booktitle={Proceedings of the 48th International ACM SIGIR Conference on Research and Development in Information Retrieval},
  pages={1934--1943},
  year={2025},
  publisher={ACM},
  doi={10.1145/3726302.3729981}
}

@inproceedings{cai2025agentic,
  title={Agentic Feedback Loop Modeling Improves Recommendation and User Simulation},
  author={Cai, Shihao and Zhang, Jizhi and Bao, Keqin and Gao, Chongming and Wang, Qifan and Feng, Fuli and He, Xiangnan},
  booktitle={Proceedings of the 48th International ACM SIGIR Conference on Research and Development in Information Retrieval},
  pages={2235--2244},
  year={2025},
  publisher={ACM},
  doi={10.1145/3726302.3729893}
}

@inproceedings{wang2025tunable,
  title={Tunable {LLM}-based Proactive Recommendation Agent},
  author={Wang, Mingze and Gao, Chongming and Wang, Wenjie and Li, Yangyang and Feng, Fuli},
  booktitle={Proceedings of the 63rd Annual Meeting of the Association for Computational Linguistics (Volume 1: Long Papers)},
  pages={19262--19276},
  year={2025},
  address={Vienna, Austria},
  publisher={Association for Computational Linguistics},
  doi={10.18653/v1/2025.acl-long.944}
}

@inproceedings{chen2025dlcrec,
  title={{DLCRec}: A Novel Approach for Managing Diversity in {LLM}-Based Recommender Systems},
  author={Chen, Jiaju and Gao, Chongming and Yuan, Shuai and Liu, Shuchang and Cai, Qingpeng and Jiang, Peng},
  booktitle={Proceedings of the Eighteenth ACM International Conference on Web Search and Data Mining},
  pages={857--865},
  year={2025},
  publisher={ACM},
  doi={10.1145/3701551.3703572}
}

@article{tiger,
  title={Recommender systems with generative retrieval},
  author={Rajput, Shashank and Mehta, Nikhil and Singh, Anima and Hulikal Keshavan, Raghunandan and Vu, Trung and Heldt, Lukasz and Hong, Lichan and Tay, Yi and Tran, Vinh and Samost, Jonah and others},
  journal={Advances in Neural Information Processing Systems},
  volume={36},
  pages={10299--10315},
  year={2023}
}

@article{minionerec,
  title={Minionerec: An open-source framework for scaling generative recommendation},
  author={Kong, Xiaoyu and Sheng, Leheng and Tan, Junfei and Chen, Yuxin and Wu, Jiancan and Zhang, An and Wang, Xiang and He, Xiangnan},
  journal={arXiv preprint arXiv:2510.24431},
  year={2025}
}

@article{rearec,
  title={Think before recommend: Unleashing the latent reasoning power for sequential recommendation},
  author={Tang, Jiakai and Dai, Sunhao and Shi, Teng and Xu, Jun and Chen, Xu and Chen, Wen and Wu, Jian and Jiang, Yuning},
  journal={arXiv preprint arXiv:2503.22675},
  year={2025}
}

@inproceedings{rrec,
  title={$\text {R}^ 2\text {ec} $: Towards Large Recommender Models with Reasoning},
  author={You, Runyang and Li, Yongqi and Lin, Xinyu and Zhang, Xin and Wang, Wenjie and Li, Wenjie and Nie, Liqiang},
  booktitle={The Thirty-ninth Annual Conference on Neural Information Processing Systems},
  year={2025}
}

@inproceedings{flame,
  title={Fine-grained List-wise Alignment for Generative Medication Recommendation},
  author={Fan, Chenxiao and Gao, Chongming and Shi, Wentao and Gong, Yaxin and Zhao, Zihao and Feng, Fuli},
  booktitle={Advances in Neural Information Processing Systems},
  volume={38},
  year={2025},
  url={https://openreview.net/forum?id=Quo3XadYcZ}
}

@inproceedings{lcrec,
  title={Adapting large language models by integrating collaborative semantics for recommendation},
  author={Zheng, Bowen and Hou, Yupeng and Lu, Hongyu and Chen, Yu and Zhao, Wayne Xin and Chen, Ming and Wen, Ji-Rong},
  booktitle={2024 IEEE 40th International Conference on Data Engineering (ICDE)},
  pages={1435--1448},
  year={2024},
  organization={IEEE}
}

@inproceedings{letter,
  title={Learnable item tokenization for generative recommendation},
  author={Wang, Wenjie and Bao, Honghui and Lin, Xinyu and Zhang, Jizhi and Li, Yongqi and Feng, Fuli and Ng, See-Kiong and Chua, Tat-Seng},
  booktitle={Proceedings of the 33rd ACM International Conference on Information and Knowledge Management},
  pages={2400--2409},
  year={2024}
}

@inproceedings{tallrec,
  title={Tallrec: An effective and efficient tuning framework to align large language model with recommendation},
  author={Bao, Keqin and Zhang, Jizhi and Zhang, Yang and Wang, Wenjie and Feng, Fuli and He, Xiangnan},
  booktitle={Proceedings of the 17th ACM conference on recommender systems},
  pages={1007--1014},
  year={2023}
}

@article{rere,
  title={Reinforced preference optimization for recommendation},
  author={Tan, Junfei and Chen, Yuxin and Zhang, An and Jiang, Junguang and Liu, Bin and Xu, Ziru and Zhu, Han and Xu, Jian and Zheng, Bo and Wang, Xiang},
  journal={arXiv preprint arXiv:2510.12211},
  year={2025}
}

@inproceedings{caser,
  title={Personalized top-n sequential recommendation via convolutional sequence embedding},
  author={Tang, Jiaxi and Wang, Ke},
  booktitle={Proceedings of the eleventh ACM international conference on web search and data mining},
  pages={565--573},
  year={2018}
}

@article{grpo,
  title={Deepseekmath: Pushing the limits of mathematical reasoning in open language models},
  author={Shao, Zhihong and Wang, Peiyi and Zhu, Qihao and Xu, Runxin and Song, Junxiao and Bi, Xiao and Zhang, Haowei and Zhang, Mingchuan and Li, YK and Wu, Yang and others},
  journal={arXiv preprint arXiv:2402.03300},
  year={2024}
}

@article{dpo,
  title={Direct preference optimization: Your language model is secretly a reward model},
  author={Rafailov, Rafael and Sharma, Archit and Mitchell, Eric and Manning, Christopher D and Ermon, Stefano and Finn, Chelsea},
  journal={Advances in neural information processing systems},
  volume={36},
  pages={53728--53741},
  year={2023}
}

@article{ppo,
  title={Proximal policy optimization algorithms},
  author={Schulman, John and Wolski, Filip and Dhariwal, Prafulla and Radford, Alec and Klimov, Oleg},
  journal={arXiv preprint arXiv:1707.06347},
  year={2017}
}

@article{rlhf,
  title={Deep reinforcement learning from human preferences},
  author={Christiano, Paul F and Leike, Jan and Brown, Tom and Martic, Miljan and Legg, Shane and Amodei, Dario},
  journal={Advances in neural information processing systems},
  volume={30},
  year={2017}
}

@inproceedings{code_calibration,
  title={Calibration and Correctness of Language Models for Code},
  author={Spiess, Claudio and Gros, David and Pai, Kunal Suresh and Pradel, Michael and Rabin, Md Rafiqul Islam and Alipour, Amin and Jha, Susmit and Devanbu, Prem and Ahmed, Toufique},
  booktitle={Proceedings of the IEEE/ACM 47th International Conference on Software Engineering},
  pages={540--552},
  numpages={13},
  year={2025},
  isbn={9798331505691},
  publisher={IEEE Press},
  doi={10.1109/ICSE55347.2025.00040},
  url={https://doi.org/10.1109/ICSE55347.2025.00040},
  location={Ottawa, Ontario, Canada},
  series={ICSE '25}
}

@inproceedings{rlcr,
  title={Beyond Binary Rewards: Training {LMs} to Reason About Their Uncertainty},
  author={Damani, Mehul and Puri, Isha and Slocum, Stewart and Shenfeld, Idan and Choshen, Leshem and Kim, Yoon and Andreas, Jacob},
  booktitle={International Conference on Learning Representations},
  year={2026},
  url={https://openreview.net/forum?id=ASQ649zdHm}
}

@inproceedings{he2017neural,
  title={Neural collaborative filtering},
  author={He, Xiangnan and Liao, Lizi and Zhang, Hanwang and Nie, Liqiang and Hu, Xia and Chua, Tat-Seng},
  booktitle={Proceedings of the 26th international conference on world wide web},
  pages={173--182},
  year={2017}
}

@inproceedings{geng2022recommendation,
  title={Recommendation as language processing (rlp): A unified pretrain, personalized prompt \& predict paradigm (p5)},
  author={Geng, Shijie and Liu, Shuchang and Fu, Zuohui and Ge, Yingqiang and Zhang, Yongfeng},
  booktitle={Proceedings of the 16th ACM conference on recommender systems},
  pages={299--315},
  year={2022}
}

@inproceedings{hou2025generative,
  title={Generative Recommendation Models: Progress and Directions},
  author={Hou, Yupeng and Zhang, An and Sheng, Leheng and Yang, Zhengyi and Wang, Xiang and Chua, Tat-Seng and McAuley, Julian},
  booktitle={Companion Proceedings of the ACM on Web Conference 2025},
  pages={13--16},
  year={2025}
}

@article{li2025survey,
  title={A survey of generative recommendation from a tri-decoupled perspective: Tokenization, architecture, and optimization},
  author={Li, Xiaopeng and Chen, Bo and She, Junda and Cao, Shiteng and Wang, You and Jia, Qinlin and He, Haiying and Zhou, Zheli and Liu, Zhao and Liu, Ji and others},
  year={2025},
  publisher={Preprints}
}

@inproceedings{wang2025generative,
  title={Generative Recommendation: Towards Personalized Multimodal Content Generation},
  author={Wang, Wenjie and Lin, Xinyu and Feng, Fuli and He, Xiangnan and Chua, Tat-Seng},
  booktitle={Companion Proceedings of the ACM on Web Conference 2025},
  pages={2421--2425},
  year={2025}
}

@inproceedings{bao2024decoding,
  title={Decoding matters: Addressing amplification bias and homogeneity issue in recommendations for large language models},
  author={Bao, Keqin and Zhang, Jizhi and Zhang, Yang and Huo, Xinyue and Chen, Chong and Feng, Fuli},
  booktitle={Proceedings of the 2024 Conference on Empirical Methods in Natural Language Processing},
  pages={10540--10552},
  year={2024}
}

@inproceedings{lin2025order,
  title={Order-agnostic identifier for large language model-based generative recommendation},
  author={Lin, Xinyu and Shi, Haihan and Wang, Wenjie and Feng, Fuli and Wang, Qifan and Ng, See-Kiong and Chua, Tat-Seng},
  booktitle={Proceedings of the 48th international ACM SIGIR conference on research and development in information retrieval},
  pages={1923--1933},
  year={2025}
}

@article{liao2024rosepo,
  title={Rosepo: Aligning llm-based recommenders with human values},
  author={Liao, Jiayi and He, Xiangnan and Xie, Ruobing and Wu, Jiancan and Yuan, Yancheng and Sun, Xingwu and Kang, Zhanhui and Wang, Xiang},
  journal={arXiv preprint arXiv:2410.12519},
  year={2024}
}

@inproceedings{cai2025mgfrec,
  title={{MGFRec}: Towards Reinforced Reasoning Recommendation with Multiple Groundings and Feedback},
  author={Cai, Shihao and Gao, Chongming and Liu, Haoyan and Shi, Wentao and Sun, Jianshan and Tang, Ruiming and Feng, Fuli},
  booktitle={Proceedings of the 32nd ACM SIGKDD Conference on Knowledge Discovery and Data Mining V.1},
  pages={49--58},
  year={2026},
  publisher={ACM},
  doi={10.1145/3770854.3780206}
}

@inproceedings{cai2025korder,
  title={K-order Ranking Preference Optimization for Large Language Models},
  author={Cai, Shihao and Gao, Chongming and Zhang, Yang and Shi, Wentao and Zhang, Jizhi and Bao, Keqin and Wang, Qifan and Feng, Fuli},
  booktitle={Findings of the Association for Computational Linguistics: ACL 2025},
  pages={4844--4859},
  year={2025},
  address={Vienna, Austria},
  publisher={Association for Computational Linguistics},
  doi={10.18653/v1/2025.findings-acl.250}
}

@inproceedings{zhang2025unpaired,
  title={Leveraging Unpaired Feedback for Long-Term {LLM}-based Recommendation Tuning},
  author={Zhang, Jizhi and Gao, Chongming and Shi, Wentao and Chen, Xin and Wang, Jingang and Cai, Xunliang and Feng, Fuli},
  booktitle={Findings of the Association for Computational Linguistics: EMNLP 2025},
  pages={24507--24521},
  year={2025},
  address={Suzhou, China},
  publisher={Association for Computational Linguistics},
  doi={10.18653/v1/2025.findings-emnlp.1332},
  url={https://aclanthology.org/2025.findings-emnlp.1332/}
}

@article{mao2025reinforcedprompt,
  title={Reinforced Prompt Personalization for Recommendation with Large Language Models},
  author={Mao, Wenyu and Wu, Jiancan and Chen, Weijian and Gao, Chongming and Wang, Xiang and He, Xiangnan},
  journal={ACM Transactions on Information Systems},
  volume={43},
  number={3},
  pages={72:1--72:27},
  year={2025},
  doi={10.1145/3716320}
}

@article{zhang2025vague,
  title={Vague Preference Policy Learning for Conversational Recommendation},
  author={Zhang, Gangyi and Gao, Chongming and Lei, Wenqiang and Guo, Xiaojie and Li, Shijun and Chen, Hongshen and Ding, Zhuozhi and Xu, Sulong and Wu, Lingfei},
  journal={ACM Transactions on Information Systems},
  volume={43},
  number={3},
  pages={78:1--78:27},
  year={2025},
  doi={10.1145/3717831}
}

@inproceedings{chen2026blade,
  title={Beyond Static Best-of-N: Bayesian List-wise Alignment for {LLM}-based Recommendation},
  author={Chen, Ruijun and Gao, Chongming and Chen, Jiawei and Yang, Weiqin and He, Xiangnan},
  booktitle={Proceedings of the 49th International ACM SIGIR Conference on Research and Development in Information Retrieval},
  year={2026},
  note={To appear},
  eprint={2605.04559},
  archivePrefix={arXiv},
  primaryClass={cs.IR}
}

@article{chen2024softmax,
  title={On softmax direct preference optimization for recommendation},
  author={Chen, Yuxin and Tan, Junfei and Zhang, An and Yang, Zhengyi and Sheng, Leheng and Zhang, Enzhi and Wang, Xiang and Chua, Tat-Seng},
  journal={Advances in Neural Information Processing Systems},
  volume={37},
  pages={27463--27489},
  year={2024}
}

@inproceedings{guo2017calibration,
  title={On calibration of modern neural networks},
  author={Guo, Chuan and Pleiss, Geoff and Sun, Yu and Weinberger, Kilian Q},
  booktitle={International conference on machine learning},
  pages={1321--1330},
  year={2017},
  organization={PMLR}
}

@article{kadavath2022language,
  title={Language models (mostly) know what they know},
  author={Kadavath, Saurav and Conerly, Tom and Askell, Amanda and Henighan, Tom and Drain, Dawn and Perez, Ethan and Schiefer, Nicholas and Hatfield-Dodds, Zac and DasSarma, Nova and Tran-Johnson, Eli and others},
  journal={arXiv preprint arXiv:2207.05221},
  year={2022}
}

@article{kendall2017uncertainties,
  title={What uncertainties do we need in bayesian deep learning for computer vision?},
  author={Kendall, Alex and Gal, Yarin},
  journal={Advances in neural information processing systems},
  volume={30},
  year={2017}
}

@inproceedings{xiao2019quantifying,
  title={Quantifying uncertainties in natural language processing tasks},
  author={Xiao, Yijun and Wang, William Yang},
  booktitle={Proceedings of the AAAI conference on artificial intelligence},
  volume={33},
  number={01},
  pages={7322--7329},
  year={2019}
}

@article{lin2022teaching,
  title={Teaching models to express their uncertainty in words},
  author={Lin, Stephanie and Hilton, Jacob and Evans, Owain},
  journal={Transactions on Machine Learning Research},
  year={2022},
  url={https://openreview.net/forum?id=8s8K2UZGTZ}
}

@inproceedings{lin2017focal,
  title={Focal loss for dense object detection},
  author={Lin, Tsung-Yi and Goyal, Priya and Girshick, Ross and He, Kaiming and Doll{\'a}r, Piotr},
  booktitle={Proceedings of the IEEE international conference on computer vision},
  pages={2980--2988},
  year={2017}
}

@article{qwen3,
  title={Qwen3 technical report},
  author={Yang, An and Li, Anfeng and Yang, Baosong and Zhang, Beichen and Hui, Binyuan and Zheng, Bo and Yu, Bowen and Gao, Chang and Huang, Chengen and Lv, Chenxu and others},
  journal={arXiv preprint arXiv:2505.09388},
  year={2025}
}

@article{wu2024survey,
  title={A survey on large language models for recommendation},
  author={Wu, Likang and Zheng, Zhi and Qiu, Zhaopeng and Wang, Hao and Gu, Hongchao and Shen, Tingjia and Qin, Chuan and Zhu, Chen and Zhu, Hengshu and Liu, Qi and others},
  journal={World Wide Web},
  volume={27},
  number={5},
  pages={60},
  year={2024},
  publisher={Springer}
}

@article{zhang2025collm,
  title={Collm: Integrating collaborative embeddings into large language models for recommendation},
  author={Zhang, Yang and Feng, Fuli and Zhang, Jizhi and Bao, Keqin and Wang, Qifan and He, Xiangnan},
  journal={IEEE Transactions on Knowledge and Data Engineering},
  year={2025},
  publisher={IEEE}
}

@article{wang2023enhancing,
  title={Enhancing recommender systems with large language model reasoning graphs},
  author={Wang, Yan and Chu, Zhixuan and Ouyang, Xin and Wang, Simeng and Hao, Hongyan and Shen, Yue and Gu, Jinjie and Xue, Siqiao and Zhang, James Y and Cui, Qing and others},
  journal={arXiv preprint arXiv:2308.10835},
  year={2023}
}

@inproceedings{bai2024aligning,
  title={Aligning large language model with direct multi-preference optimization for recommendation},
  author={Bai, Zhuoxi and Wu, Ning and Cai, Fengyu and Zhu, Xinyi and Xiong, Yun},
  booktitle={Proceedings of the 33rd ACM International Conference on Information and Knowledge Management},
  pages={76--86},
  year={2024}
}

@inproceedings{kong2025think,
  title={Think before Recommendation: Autonomous Reasoning-enhanced Recommender},
  author={Kong, Xiaoyu and Jiang, Junguang and Liu, Bin and Xu, Ziru and Zhu, Han and Xu, Jian and Zheng, Bo and Wu, Jiancan and Wang, Xiang},
  booktitle={Advances in Neural Information Processing Systems},
  volume={38},
  year={2025},
  url={https://papers.nips.cc/paper_files/paper/2025/hash/cea5bc68b890bffb10f18aaaab2becb1-Abstract-Conference.html}
}

@article{lin2025rec,
  title={Rec-r1: Bridging generative large language models and user-centric recommendation systems via reinforcement learning},
  author={Lin, Jiacheng and Wang, Tian and Qian, Kun},
  journal={Transactions on Machine Learning Research},
  year={2025},
  url={https://openreview.net/forum?id=YBRU9MV2vE}
}

@article{ding2026towards,
  title={Towards Sample-Efficient and Stable Reinforcement Learning for LLM-based Recommendation},
  author={Ding, Hongxun and Bao, Keqin and Zhang, Jizhi and Fang, Yi and Xu, Wenxin and Feng, Fuli and He, Xiangnan},
  journal={arXiv preprint arXiv:2602.00632},
  year={2026}
}

@article{huang2025survey,
  title={A survey on hallucination in large language models: Principles, taxonomy, challenges, and open questions},
  author={Huang, Lei and Yu, Weijiang and Ma, Weitao and Zhong, Weihong and Feng, Zhangyin and Wang, Haotian and Chen, Qianglong and Peng, Weihua and Feng, Xiaocheng and Qin, Bing and others},
  journal={ACM Transactions on Information Systems},
  volume={43},
  number={2},
  pages={1--55},
  year={2025},
  publisher={ACM New York, NY}
}

@inproceedings{tian2023just,
  title={Just Ask for Calibration: Strategies for Eliciting Calibrated Confidence Scores from Language Models Fine-Tuned with Human Feedback},
  author={Tian, Katherine and Mitchell, Eric and Zhou, Allan and Sharma, Archit and Rafailov, Rafael and Yao, Huaxiu and Finn, Chelsea and Manning, Christopher D},
  booktitle={Proceedings of the 2023 Conference on Empirical Methods in Natural Language Processing},
  pages={5433--5442},
  year={2023},
  address={Singapore},
  publisher={Association for Computational Linguistics},
  doi={10.18653/v1/2023.emnlp-main.330}
}

@inproceedings{rqvae,
  title={Autoregressive image generation using residual quantization},
  author={Lee, Doyup and Kim, Chiheon and Kim, Saehoon and Cho, Minsu and Han, Wook-Shin},
  booktitle={Proceedings of the IEEE/CVF conference on computer vision and pattern recognition},
  pages={11523--11532},
  year={2022}
}

@inproceedings{lora,
  title={{LoRA}: Low-Rank Adaptation of Large Language Models},
  author={Hu, Edward J and Shen, Yelong and Wallis, Phillip and Allen-Zhu, Zeyuan and Li, Yuanzhi and Wang, Shean and Wang, Lu and Chen, Weizhu},
  booktitle={International Conference on Learning Representations},
  year={2022},
  url={https://openreview.net/forum?id=nZeVKeeFYf9}
}

@inproceedings{kweon2025uncertainty,
  title={Uncertainty Quantification and Decomposition for LLM-based Recommendation},
  author={Kweon, Wonbin and Jang, Sanghwan and Kang, SeongKu and Yu, Hwanjo},
  booktitle={Proceedings of the ACM on Web Conference 2025},
  pages={4889--4901},
  year={2025}
}

\appendix

\section{Datasets}
\label{appendix:datasets}

We conduct experiments on three real-world benchmarks: Office Products (Office) and Industrial and Scientific (Industrial) from the Amazon Review Dataset\footnote{\url{https://jmcauley.ucsd.edu/data/amazon}}, along with the Yelp dataset\footnote{\url{https://www.yelp.com/dataset}}.
To ensure data quality and computational efficiency, we applied the following preprocessing steps:

\begin{enumerate}[leftmargin=*]
    \item \textbf{5-core Filtering:} Removing users and items with fewer than five interactions.
    \item \textbf{Time Window Selection:} Selecting interactions within specific time ranges (as detailed in Table~\ref{tab:dataset_stats}) to maintain appropriate dataset sizes and focus on relevant interaction patterns.
    \item \textbf{Sequence Truncation:} Truncating user interaction histories to a maximum sequence length of 10.
    \item \textbf{Chronological Splitting:} Partitioning data into training, validation, and test sets with an 8:1:1 ratio based on the strict chronological order of interactions.
\end{enumerate}

The detailed statistics of the processed datasets are summarized in Table~\ref{tab:dataset_stats}.

\begin{table}[h]
\centering
\caption{Statistics of the preprocessed datasets.}
\label{tab:dataset_stats}
\resizebox{\linewidth}{!}{
\begin{tabular}{l c c c c}
\toprule
\textbf{Dataset} & \textbf{Time Range} & \textbf{\# Users} & \textbf{\# Items} & \textbf{Train/Valid/Test} \\
\midrule
Office & 2016.10 -- 2018.11 & 8,319 & 3,452 & 38,883 / 4,860 / 4,861 \\
Industrial & 1996.10 -- 2018.11 & 7,702 & 3,689 & 36,294 / 4,537 / 4,537 \\
Yelp & 2018.05 -- 2018.11 & 6,231 & 4,852 & 37,579 / 4,697 / 4,698 \\
\bottomrule
\end{tabular}
}
\end{table}

\section{Baselines}
\label{appendix:baselines}
We select a comprehensive set of baselines for comparison.
All baselines are implemented via official repositories, with the best validation checkpoints selected for evaluation.
The baselines are categorized into three groups: traditional recommendation models, LLM-based generative models (subdivided into text-based methods and SID-based generative methods), and reasoning-enhanced models.

\vspace{5pt}
\noindent \textbf{1. Traditional Recommendation Models}
\begin{itemize}[leftmargin=*]
    \item \textbf{SASRec}~\cite{sasrec} is a sequential recommendation model equipped with self-attention mechanisms to capture long-term semantic dependencies.
\end{itemize}

\vspace{5pt}
\noindent \textbf{2. LLM-based Generative Models}
We further divide this category into two streams based on their generation targets:

\noindent \textit{(1) Text-based Methods:}
\begin{itemize}[leftmargin=*]
    \item \textbf{BIGRec}~\cite{bigrec} proposes a bi-step grounding paradigm that fine-tunes LLMs to generate meaningful semantic tokens, which are subsequently mapped to actual items via a grounding mechanism.
    \item \textbf{SPRec}~\cite{sprec} is a self-play framework that iteratively aligns the model via DPO using self-generated negative samples to mitigate the filter bubble effect.
\end{itemize}

\noindent \textit{(2) SID-based Generative Methods:}
\begin{itemize}[leftmargin=*]
    \item \textbf{TIGER}~\cite{tiger} is a transformer-based model that utilizes RQ-VAE to acquire item SIDs for autoregressive generation.
    \item \textbf{MiniOneRec}~\cite{minionerec} is an open-source generative framework that establishes an end-to-end workflow, utilizing RQ-VAE for item indexing and employing reinforcement learning with hybrid rewards for optimization.
\end{itemize}

\vspace{5pt}
\noindent \textbf{3. Reasoning-enhanced Models}
\begin{itemize}[leftmargin=*]
    \item \textbf{ReaRec}~\cite{rearec} is an inference-time computing framework that enhances user representations through implicit autoregressive multi-step reasoning.
    \item \textbf{R$^2$ec}~\cite{rrec} is a unified framework featuring a dual-head architecture that jointly optimizes reasoning chain generation and item prediction.
\end{itemize}

\section{Implementation Details}
\label{appendix:implementation}

\noindent\textbf{Model Architecture and Pre-training.}
We employ Qwen3-8B~\cite{qwen3} as the backbone LLM and represent items using 4-layer SIDs. The item embeddings are initialized via Qwen3-Embedding-4B, and the underlying RQ-VAE~\cite{rqvae} is trained on single GPU using the AdamW optimizer with a learning rate of 1e-3 and a batch size of 1,024 for 10,000 epochs.

\noindent\textbf{Training Protocol.}
The training pipeline consists of two stages performed on 8 NVIDIA A100 GPUs.
First, during the SFT stage, we apply LoRA~\cite{lora} ($r=16, \alpha=32$) and train the model for 10 epochs using a learning rate of 2e-4 with a cosine decay schedule and a per-device batch size of 8. We evaluate the model every 0.5 epochs with a patience of 3 evaluations for early stopping.
Subsequently, we warm-start from the SFT checkpoint to perform Uncertainty-Aware Preference Alignment. We lower the learning rate to 2e-6 while maintaining the same LoRA configuration. For the optimization objective, we set the KL penalty $\beta=0$ and generate $G=16$ candidates per sample via beam search. The hyperparameters are tuned via grid search: $\alpha \in \{0, 0.5, 0.8\}$, $\gamma_1 \in \{0.0, 0.1, 0.5\}$, and $\gamma_2 \in \{0.5, 1.0\}$.

\noindent\textbf{Inference and Prompting.}
During inference, we generate recommendation lists using beam search with a beam size of 10. The explicit confidence scores are computed post-generation via Eq.~(\ref{equ:conf}).
We format the input as a natural language instruction with special tokens for SIDs. An example is provided below:

\begin{tcolorbox}[colback=gray!5, colframe=gray!40, boxrule=0.5pt, arc=2pt, left=5pt, right=5pt, top=5pt, bottom=5pt]
\small
\textbf{Input:} \\
The user has interacted with items \texttt{<a\_147><b\_0><c\_248><d\_36>}, \texttt{<a\_147><b\_0><c\_248><d\_36>}, \texttt{<a\_73><b\_204><c\_17><d\_180>} in chronological order. Can you predict the next possible item that the user may expect?

\vspace{0.3em}
\textbf{Response:} \\
\texttt{<a\_70><b\_140><c\_195><d\_192>}
\end{tcolorbox}

\noindent\textbf{Risk-Aware Scoring Details.}
For the downstream applications in Section~\ref{rq4} (i.e., re-ranking, user rejection, and item truncation), we adopt a hybrid scoring strategy that integrates both the generative likelihood and the explicit confidence. Specifically, the final ranking score is calculated as $Score = \exp(s_{\text{gen}}) + \lambda \cdot s_{\text{conf}}$, where $s_{\text{gen}}$ denotes the sequence log-likelihood, $s_{\text{conf}}$ is the explicit confidence score, and $\lambda$ is a balancing factor set to 0.5.
For the User-level Rejection task, we represent the confidence of a user request using the maximum hybrid score among its generated candidates.

\noindent\textbf{Computational Cost.}
Table~\ref{tab:cost} compares resource consumption on the Office dataset.

\begin{table}[h]
\centering
\caption{Computational cost comparison on Office.}
\label{tab:cost}
\begin{tabular}{l ccc}
\toprule
\textbf{Method} & \textbf{GPU Mem} & \textbf{Train / Epoch} & \textbf{Inference / Query} \\
\midrule
SFT & 34G & 7 min & 32.9 ms \\
GRPO & 50G & 65 min & 32.9 ms \\
UGR & 60G & 80 min & 37.0 ms \\
\bottomrule
\end{tabular}
\end{table}

Compared with GRPO, UGR introduces +10G memory and +15 min per epoch, primarily due to the auxiliary confidence alignment loss and one additional forward pass for confidence estimation. At inference time, the extra 4.1 ms latency is negligible and enables the downstream risk-aware applications described in Section~\ref{rq4}, which standard methods cannot support.

\section{Additional Ablation Studies}
\label{appendix:ablation}

\noindent\textbf{Complete Component-wise Ablation.}
Table~\ref{tab:component_ablation} reports all seven combinations across three datasets; the first row is the SFT baseline. E alone collapses far below the baseline, confirming that confidence alignment requires the reward structure from U or D. Among two-component variants, D+E is the strongest yet still trails the full model on all metrics. Only the complete U+D+E consistently achieves the best performance, showing that the three modules provide complementary contributions.

\begin{table}[h]
\centering
\caption{Complete component-wise ablation. U: Uncertainty-Weighted Reward; D: Difficulty-aware Optimization; E: Explicit Confidence Alignment.}
\label{tab:component_ablation}
\resizebox{\linewidth}{!}{
\begin{tabular}{ccc cc cc cc}
\toprule
\multicolumn{3}{c}{\textbf{Components}} & \multicolumn{2}{c}{\textbf{Office}} & \multicolumn{2}{c}{\textbf{Industrial}} & \multicolumn{2}{c}{\textbf{Yelp}} \\
\cmidrule(lr){1-3} \cmidrule(lr){4-5} \cmidrule(lr){6-7} \cmidrule(lr){8-9}
U & D & E & HR@10 & NG@10 & HR@10 & NG@10 & HR@10 & NG@10 \\
\midrule
\multicolumn{3}{c}{SFT} & 0.1563 & 0.1213 & 0.1378 & 0.1010 & 0.0423 & 0.0217 \\
\midrule
\cmark & & & 0.1522 & 0.1193 & 0.1338 & 0.1029 & 0.0424 & 0.0222 \\
& \cmark & & 0.1549 & 0.1202 & 0.1353 & 0.1022 & 0.0450 & 0.0236 \\
& & \cmark & 0.0710 & 0.0530 & 0.0789 & 0.0616 & 0.0415 & 0.0220 \\
\midrule
\cmark & \cmark & & 0.1579 & 0.1223 & 0.1349 & 0.1030 & 0.0454 & 0.0239 \\
\cmark & & \cmark & 0.1530 & 0.1196 & 0.1363 & 0.1023 & 0.0430 & 0.0225 \\
& \cmark & \cmark & 0.1583 & 0.1216 & 0.1373 & 0.1028 & 0.0456 & 0.0240 \\
\midrule
\cmark & \cmark & \cmark & \textbf{0.1598} & \textbf{0.1231} & \textbf{0.1393} & \textbf{0.1033} & \textbf{0.0473} & \textbf{0.0250} \\
\bottomrule
\end{tabular}
}
\end{table}

\noindent\textbf{Confidence Token Design Choice.}
Table~\ref{tab:token_design} compares our binary design against multi-level discrete tokens and a regression head on Office. The binary design outperforms all alternatives. We attribute this to two factors: (i) recommendation supervision is intrinsically sparse (only one ground-truth per user), which is insufficient for fine-grained calibration; and (ii) binary tokens align naturally with GRPO's pairwise ranking as they can be directly contrasted within the group-relative advantage, whereas multi-level or regression targets introduce competing learning signals.

\begin{table}[h]
\centering
\caption{Confidence token design comparison on Office.}
\label{tab:token_design}
\resizebox{\linewidth}{!}{
\begin{tabular}{l c c c c}
\toprule
& \textbf{Binary (Ours)} & \textbf{5-Token} & \textbf{10-Token} & \textbf{Regression} \\
\midrule
HR@10   & \textbf{0.1598} & 0.1587 & 0.1579 & 0.1582 \\
NG@10   & \textbf{0.1231} & 0.1220 & 0.1220 & 0.1222 \\
\bottomrule
\end{tabular}
}
\end{table}

\end{document}